\documentclass[prd,superscriptaddress,a4paper,showpacs,showkeys,10pt,nofootinbib]{revtex4}
\usepackage{graphicx}
\usepackage{graphics}
\usepackage{epsfig}
\usepackage{dcolumn}
\usepackage{bm}
\usepackage{multirow}
\usepackage{tabularx}
\usepackage{hyperref}
\usepackage{commath}
\usepackage{epstopdf}
\usepackage[T1]{fontenc}
\usepackage{geometry}
\usepackage{color}

\def\lsim{\raise0.3ex\hbox{$<$\kern-0.75em\raise-1.1ex\hbox{$\sim$}}}

\def\gsim{\raise0.3ex\hbox{$>$\kern-0.75em\raise-1.1ex\hbox{$\sim$}}}

\newcommand{\be}{\begin{equation}}

\newcommand{\ee}{\end{equation}}

\def\beq{\begin{equation}}

\def\eeq{\end{equation}}

\def\beqa{\begin{eqnarray}}

\def\eeqa{\end{eqnarray}}

\newcommand{\ba}{\begin{eqnarray}}

\newcommand{\ea}{\end{eqnarray}}

\def\gappeq{\mathrel{\rlap {\raise.5ex\hbox{$>$}}

{\lower.5ex\hbox{$\sim$}}}}

\def\lappeq{\mathrel{\rlap{\raise.5ex\hbox{$<$}}

{\lower.5ex\hbox{$\sim$}}}}

\def\Toprel#1\over#2{\mathrel{\mathop{#2}\limits^{#1}}}

\begin{document}

\title{Looking for new strategies to probe low mass axion-like particles in ultraperipheral heavy - ion collisions at the LHC}

\author{Pedro {\sc Nogarolli}}
\email{nogarollipedro@gmail.com} 
\affiliation{
 Instituto de F\'\i sica, Universidade Federal do Rio de Janeiro,\\
 CEP 21941-909 Rio de Janeiro, RJ, Brazil 
}

\author{Victor P. {\sc Gon\c{c}alves}}
\email{barros@ufpel.edu.br}
\affiliation{Institute of Physics and Mathematics, Federal University of Pelotas, \\
  Postal Code 354,  96010-900, Pelotas, RS, Brazil}

\author{Murilo  {\sc S. Rangel}}
 \email{rangel@if.ufrj.br} 
 \affiliation{ Instituto de F\'\i sica, Universidade Federal do Rio de Janeiro,\\ CEP 21941-909 Rio de Janeiro, RJ, Brazil }


\begin{abstract}

The possibility of searching for long-lived axion-like particles (ALPs) decaying into photons is investigated  in ultraperipheral $PbPb$ collisions at the Large Hadron Collider (LHC). 
We propose a search strategy for low mass ALPs using the LHCb and ALICE experiments. The ALP identification is performed by requiring the decay vertex be reconstructed outside the region where a primary vertex is expected, which strongly suppress the contribution associated with the decay of light mesons. We also use the fact that a fraction of the photons convert into electron-positron pairs, allowing the reconstruction of the particle decay position. We present the predictions for the pseudo - rapidity and transverse momentum distributions of the ALPs and photons. Moreover, predictions for the fiducial  cross-sections, derived considering the characteristics of the ALICE and LHCb detectors, are presented for different values of the ALP mass and the ALP - photon coupling.
\end{abstract}

\maketitle

\section{Introduction}

The description of the Dark Matter (DM) is one of the main current theoretical challenges of Particle Physics. Among the compelling DM candidates are axion-like particles (ALPs), which  arise in beyond de Standard Model (BSM) theories  
as a pseudo-Nambu-Goldstone boson resulting
from the breaking of a U(1) symmetry and are expected to couple to the Standard Model (SM) fields with model-dependent couplings (For reviews see, e.g. Refs. \cite{Graham:2015ouw,Schoeffel:2020svx,dEnterria:2021ljz}).                                      In the particular case where  the ALP  couples  to the SM particles only electromagnetically,   the Lagrangian is given by
\begin{equation}
\mathcal{L} = \frac{1}{2}(\partial_{\mu}a)^{2}- \frac{1}{2}m_{a}^{2}-\frac{1}{4}g_{a\gamma\gamma}F_{\mu\nu}\tilde{F}^{\mu\nu},
\end{equation}
where $a$ is the ALP field, $m_{a}$ is the ALP mass and $g_{a\gamma\gamma}$ is the ALP - photon coupling. In recent   years, the  searching of ALPs has been performed in  $e^+e^-$, $ep$,  $pp$, $pA$ and $AA$ collisions 
(See e.g. Refs. \cite{Jaeckel:2015jla,Bauer:2017ris,knapen,Aloni:2018vki,royon,Aloni:2019ruo,Bauer:2018uxu,Yue:2019gbh,Ebadi:2019gij,Coelho:2020saz,Inan:2020aal, Klusek-Gawenda:2019ijn, Goncalves:2021pdc, Zhang:2021sio, Yue:2021iiu, RebelloTeles:2023uig, Balkin:2023gya}), and the experimental results have imposed important constraints on the allowed values for $m_a$ and   $g_{a\gamma\gamma}$ \cite{CMS:2018erd,ATLAS:2020hii}. However, the probing of ALPs with small masses  ($\le 1.0$ GeV) is still a challenge, mainly due to the large background associated with the light meson decays, which also decay into two  photons (For a more detailed discussion see, e.g., Refs. \cite{Klusek-Gawenda:2019ijn,Goncalves:2021pdc,RebelloTeles:2023uig}).  Such studies indicate that a new search strategy is needed in order to separate the signal  from the background  in this kinematical region. One possible strategy has been proposed 
in Ref.~\cite{Alonso-Alvarez:2023wni}, which have considered  the identification of photons converted into electron-positron pairs inside the tracking detector as a way to 
improve the signal / background ratio in the searching of ALPs produced in $pp$ collisions. Our goal in this paper is to extend this proposal for ultraperipheral heavy ion collisions (UPCs) \cite{Baltz:2007kq}, using the fact that in this case the decay vertex can be selected to be outside the region where a primary vertex is expected (luminosity region), and estimate the associated cross-sections. Such requirement strongly suppresses the background contribution, since in this case the primary vertex occurs inside the luminosity region. As we will focus on low transverse energy photons, this study will provide predictions for the ALICE and LHCb experiments, which are the current detectors able to probe these photons.

This paper is organized as follows. In the next section, we present a brief review of the formalism needed to describe the ALP production in UPCs and describe the search strategy considered in our analysis. In section~\ref{sec:res}, we will present initially our predictions at the generator level for the pseudorapidity and transverse momentum distributions of an ALP  produced in ultraperipheral $PbPb$ collisions at $\sqrt{s_{NN}} = 5.5$ TeV. The associated distributions for the photons resulting from ALP decays will also be presented. In addition, the results for the fiducial cross-sections, derived by considering the kinematical range covered by the ALICE and LHCb detectors and the implementation of the proposed search strategy, will be shown and discussed. Finally, in section~\ref{sec:sum}, our main conclusions will be summarized.

\begin{figure}[t]
        \includegraphics[width=0.6\textwidth]{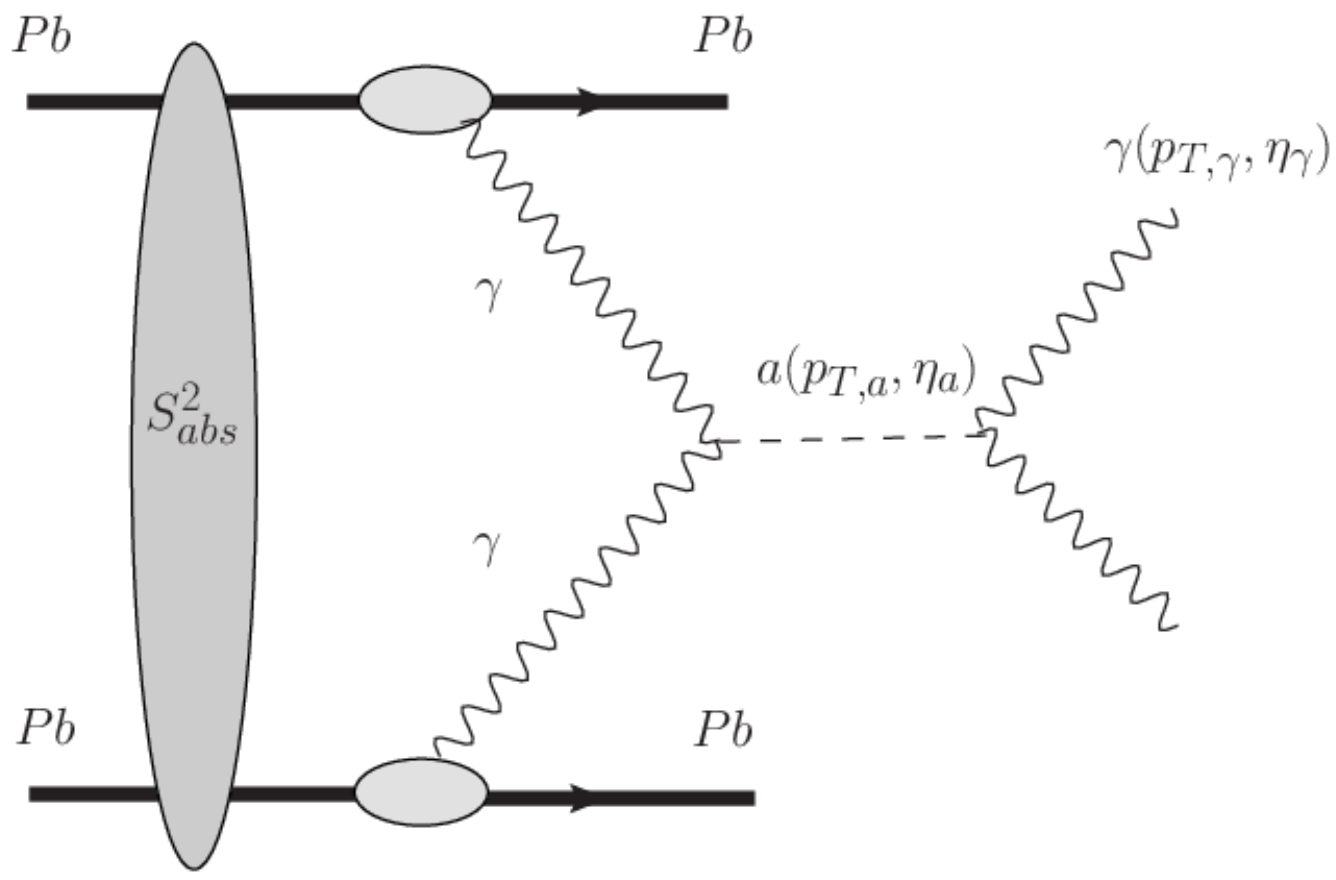}
        \caption{ALP production and decay into a two - photon system through the $\gamma \gamma \rightarrow a \rightarrow\gamma \gamma$ subprocess  in $PbPb$ collisions. }
         \label{fig:diagram}
\end{figure}

\section{Search strategy}

The production of axion - like particles in UPCs and its decay into a two-photon system is  represented in Fig. \ref{fig:diagram}. The associated cross - section  can be estimated using  the equivalent photon approximation \cite{Budnev:1975poe,Baltz:2007kq}, which implies that the total cross-section can be factorized as follows
\begin{eqnarray}
\sigma \left(Pb Pb \rightarrow Pb \otimes \gamma \gamma \otimes Pb;s_{NN} \right)   
&=& \int \mbox{d}^{2} {\mathbf r_{1}}
\mbox{d}^{2} {\mathbf r_{2}} 
\mbox{d}W 
\mbox{d}y \frac{W}{2} \, \hat{\sigma}\left(\gamma \gamma \rightarrow a \rightarrow \gamma \gamma ; 
W \right )  N\left(\omega_{1},{\mathbf r_{1}}  \right )
 N\left(\omega_{2},{\mathbf r_{2}}  \right ) S^2_{abs}({\mathbf b})  
  \,\,,
\label{cross-sec-2}
\end{eqnarray}
where $N(\omega_i, {\mathbf r}_i)$ is the number the photons  with energy $\omega_i$ at a transverse distance ${\mathbf r}_i$  from the center of nucleus, defined in the plane transverse to the trajectory, which is determined by the charge form factor of the nucleus (For a more detailed discussion see, e.g. Ref. \cite{Azevedo:2019fyz}). 
Moreover, $\hat{\sigma}$ is the cross-section of the  $\gamma \gamma \rightarrow a \rightarrow \gamma \gamma$ process,  $\sqrt{s_{NN}}$ is the center - of - mass energy of the $PbPb$ collision, $\otimes$ characterizes a rapidity gap in the final state and
$W = \sqrt{4 \omega_1 \omega_2}$ is the invariant mass of the $\gamma \gamma$ system. Finally, in order to exclude the overlap between the colliding nuclei and insure the dominance of the electromagnetic interaction, it is useful to include in Eq.(\ref{cross-sec-2}) the absorptive factor $S^2_{abs}({\mathbf b})$, which depends on the impact parameter ${\mathbf b}$ of the $PbPb$ collision.  One has that  the resulting  final state is very clean, consisting  of the diphoton system,  two intact nuclei and  two rapidity gaps, i.e. empty regions  in pseudo-rapidity that separate the intact very forward nuclei from the $\gamma \gamma$ system.
For heavy-ion collisions, the large photon-photon luminosity ($\propto Z_1^2Z_2^2$, where $Z_i$ are the atomic numbers of the incident particles) implies a large enhancement of the two - photon cross - sections, which allows probing the production of rare SM processes, as e.g. the light - by - light scattering, as well the searching of new physics such as ALPs coupled to photons \cite{Schoeffel:2020svx}. In recent years, ATLAS and CMS Collaborations reported limits on the  ALPs properties considering the  diphoton production in UPCs, providing the strongest limits to date in the properties in the mass region 5~GeV~$\leq m_a \leq$~100~GeV~\cite{CMS:2018erd,ATLAS:2020hii}.

\begin{figure}[t]
    \centering   \includegraphics[width=1.0\textwidth]{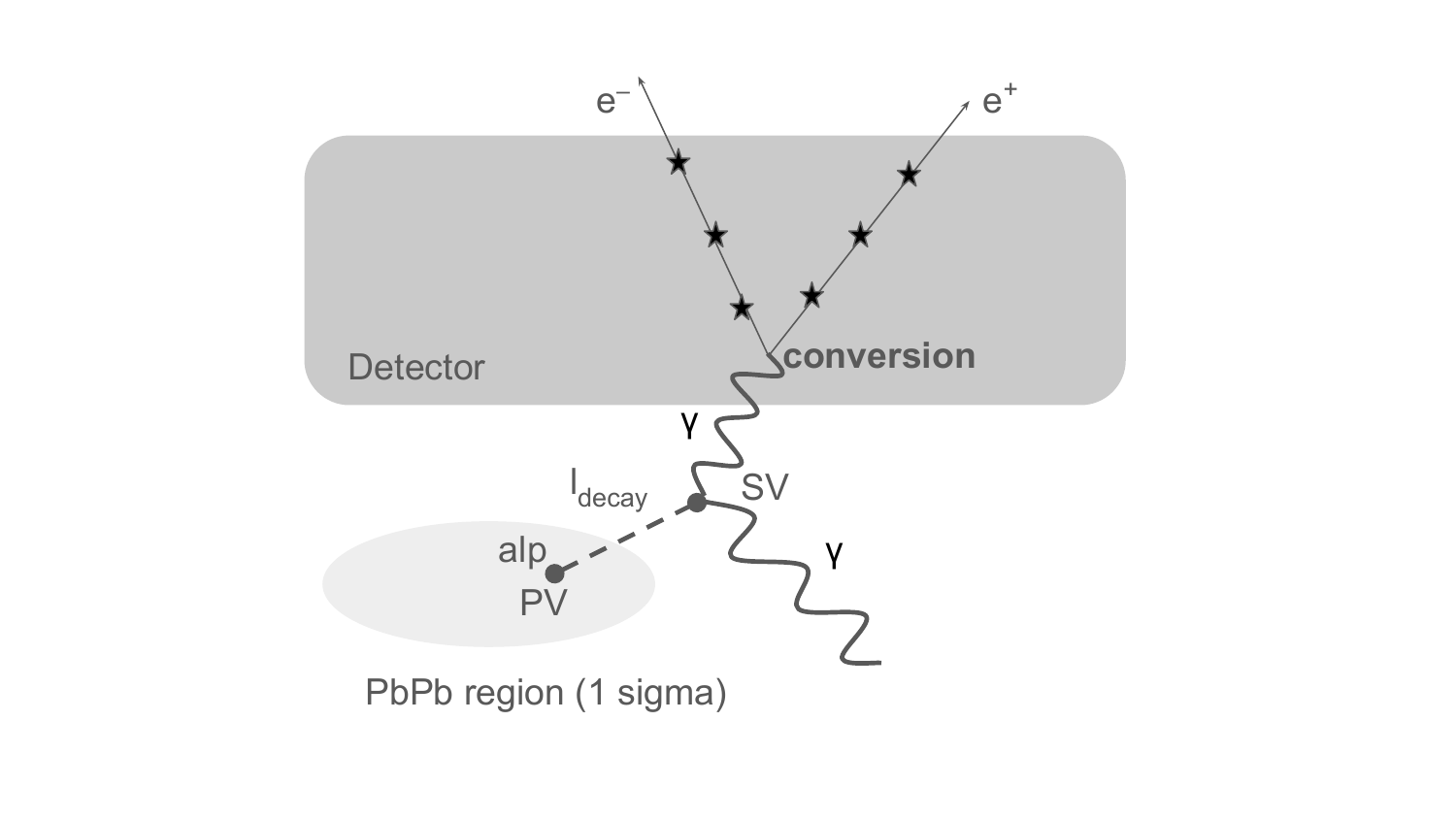}        
    \caption{Diagram showing the transverse plane of an ALP decaying into  photons outside the luminous region of the PbPb collision. Since the ALP length decay $l_{decay}$ is greater than expected PbPb collision primary vertex (PV), the secondary vertex (SV) of the ALP decay can be identified using photons converted in electron-positron pairs inside the detector.}
    \label{searchStrategy}
\end{figure}

As discussed in the Introduction, our focus in this paper is on ALPs with small masses, which generate a diphoton system in its decay. A shortcoming to separate the ALP signal in this region is associated with the fact that the diphoton low mass spectrum  has several Standard Model background sources such as $\pi^{0}\pi^{0}$, $\eta$ and $\eta^{,}(958)$ \cite{Klusek-Gawenda:2019ijn}. 
Given the values for the  cross - section associated with the ALP production, these sources of background obscure the ALP signal in this kinematical region. 
To address this challenge and establish a distinct signal region, one promising strategy involves probing ALP  that decay into photons outside the primary vertex luminous region (PV), followed by their conversion into an electron-positron pair. Fig.~\ref{searchStrategy} depicts this search strategy. The basic idea is that the reconstructed converted photons can subsequently identify a displaced secondary vertex situated beyond the lead-lead luminous region, differently from the SM light mesons that decay inside the luminous region.
 One has that the decay rate of an ALP into two photons is determined by its mass and the magnitude of the ALP - photon coupling as follows \cite{Jaeckel:2015jla}
\begin{equation}
\Gamma(a\rightarrow \gamma\gamma)=\frac{g_{a\gamma\gamma}^{2}m_{a}^{3}}{64\pi}.
\end{equation}
and the ALP decay length is determined by the inverse of the decay rate. As a consequence, one has that ALPs with small masses imply, for a fixed value of $g_{a\gamma\gamma}$,  a larger decay length.  Accounting for a relativistic boost $p_{a}/m_{a}$ and  considering a ALP with $p_{a}=0.4$~GeV, $m_{a}=0.2$ GeV and $g_{a\gamma\gamma} = 0.1  \;\text{TeV}^{-1}$,  the decay length $l_{\text{decay}}$ can be written as
\begin{equation}
l_{\text{decay}} \approx 0.1 \; \text{cm} \left(\frac{p_{a}}{0.4 \; \text{GeV}}\right)\left(\frac{0.2 \; \text{GeV}}{m_{a}}\right)^{4}\left(\frac{0.1  \;\text{TeV}^{-1}}{g_{a\gamma\gamma}}\right)^{2}\,\,
\label{ldecay}
\end{equation}
which indicates that $l_{\text{decay}}$ increases for larger values of $p_a$ and smaller values of $m_a$ and $g_{a\gamma\gamma}$. On the other hand, the primary vertex luminous region is dependent on the characteristic of each detector. In our analysis, we will consider the LHCb and ALICE detectors.
The LHCb experiment is a single-arm forward spectrometer covering the pseudorapidity range $2<\eta<5$~\cite{Aaij:2014zzy}. 
Due to material interaction,
25$\%$ of the photons convert into a pair of electron-positron that are reconstructed as tracks~\cite{Benson:2019wvt}. 
In contrast, at the ALICE experiment, the same phenomena can take place with a probability of 8.5~\%~\cite{ALICE:2014sbx} for photons with $\abs{\eta}<0.9$. These photons are reconstructed as two-track candidates with invariant masses compatible with zero, and their flight direction can be determined with high precision.

\section{Results}
\label{sec:res}

\begin{figure}[t]
    \centering
    \begin{tabular}{cc}
    \includegraphics[width=.33\textwidth]{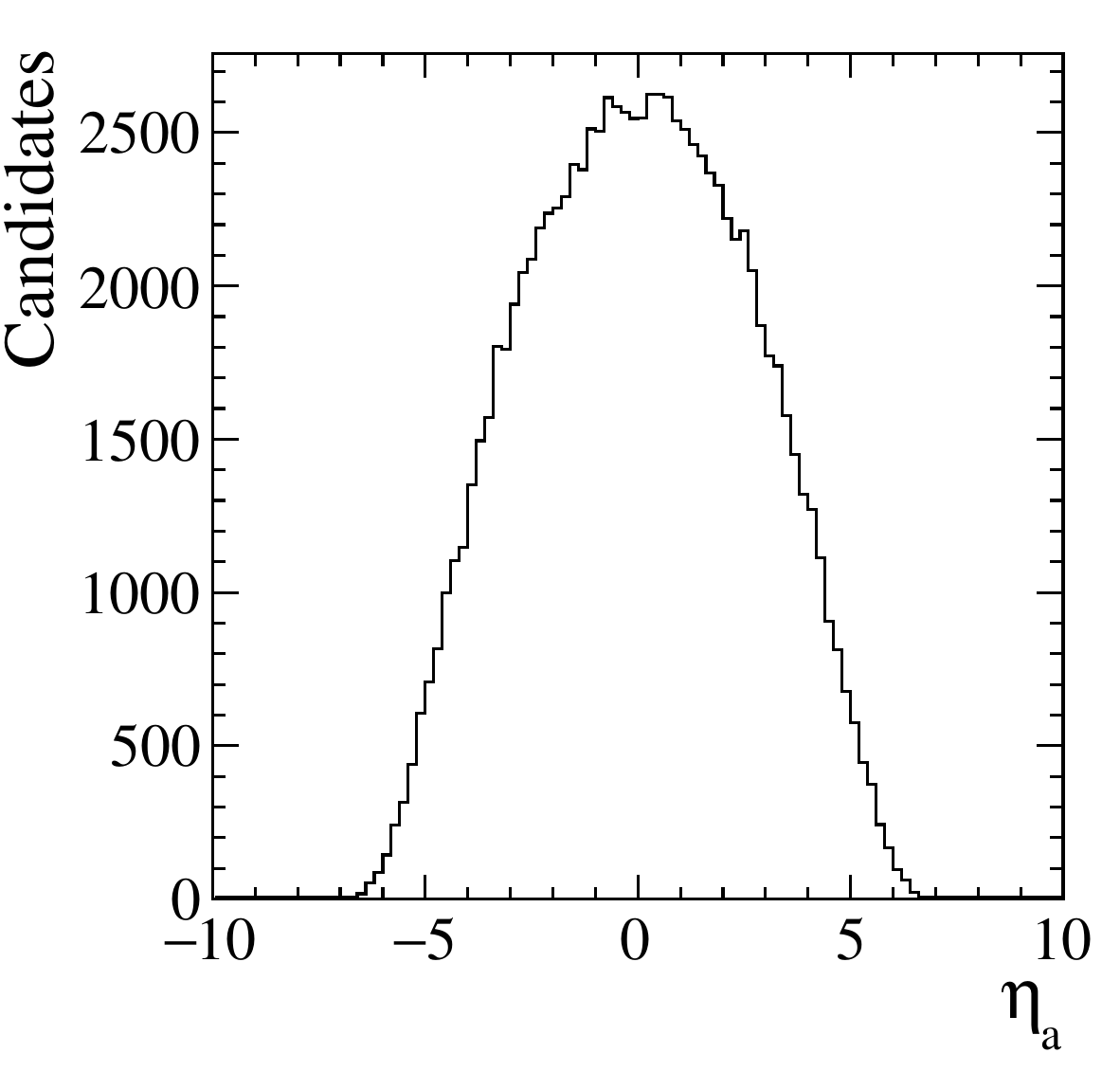} &
    \includegraphics[width=.33\textwidth]{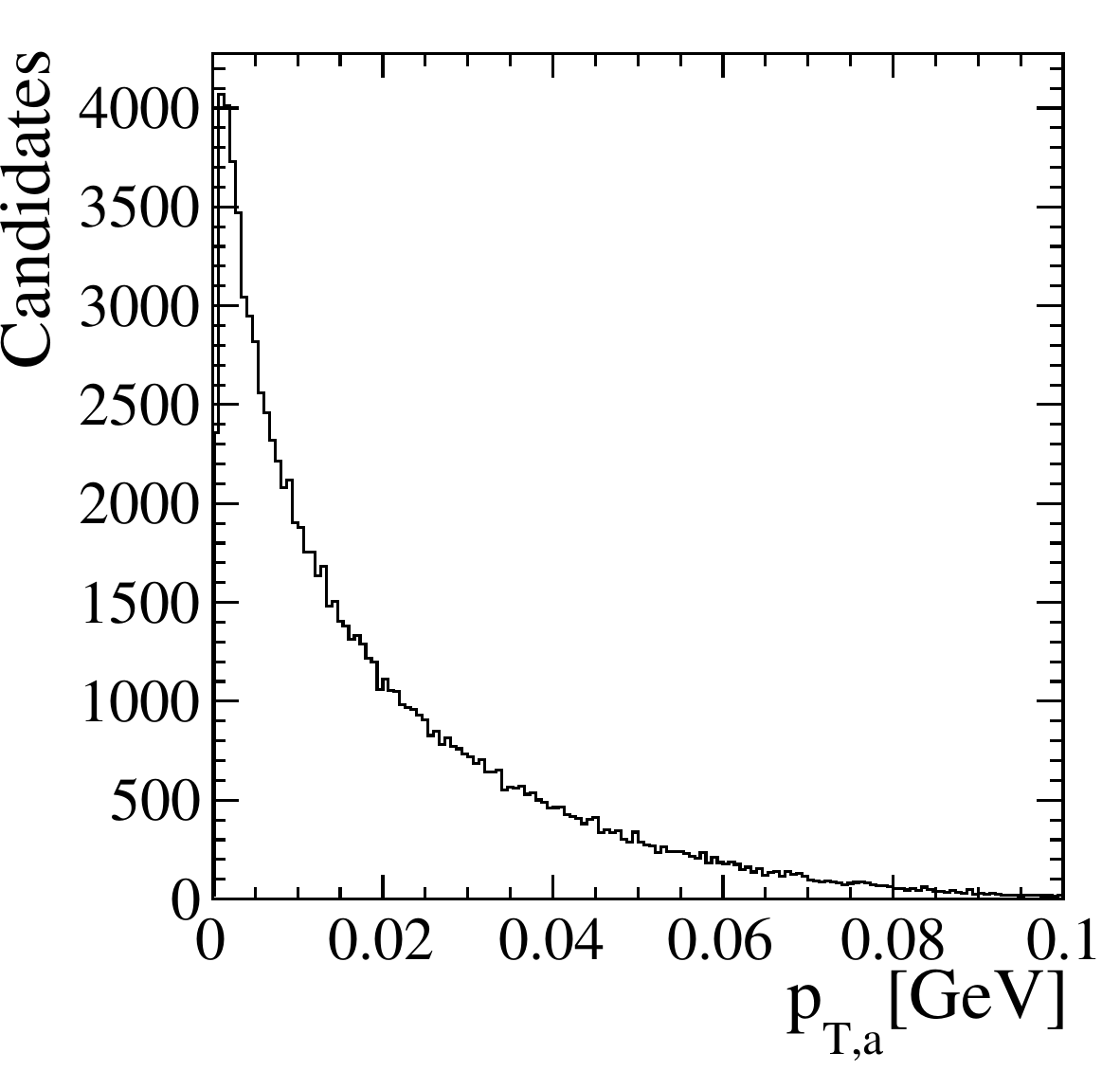}
    \end{tabular}
    \caption{Distributions of (left) pseudorapidity and (right) transverse momentum of ALP candidates with $m_a=0.5$~GeV.}
    \label{paxion}
\end{figure}

\begin{figure}[t]
    \centering
    \begin{tabular}{ccc}
    \includegraphics[width=.33\textwidth]{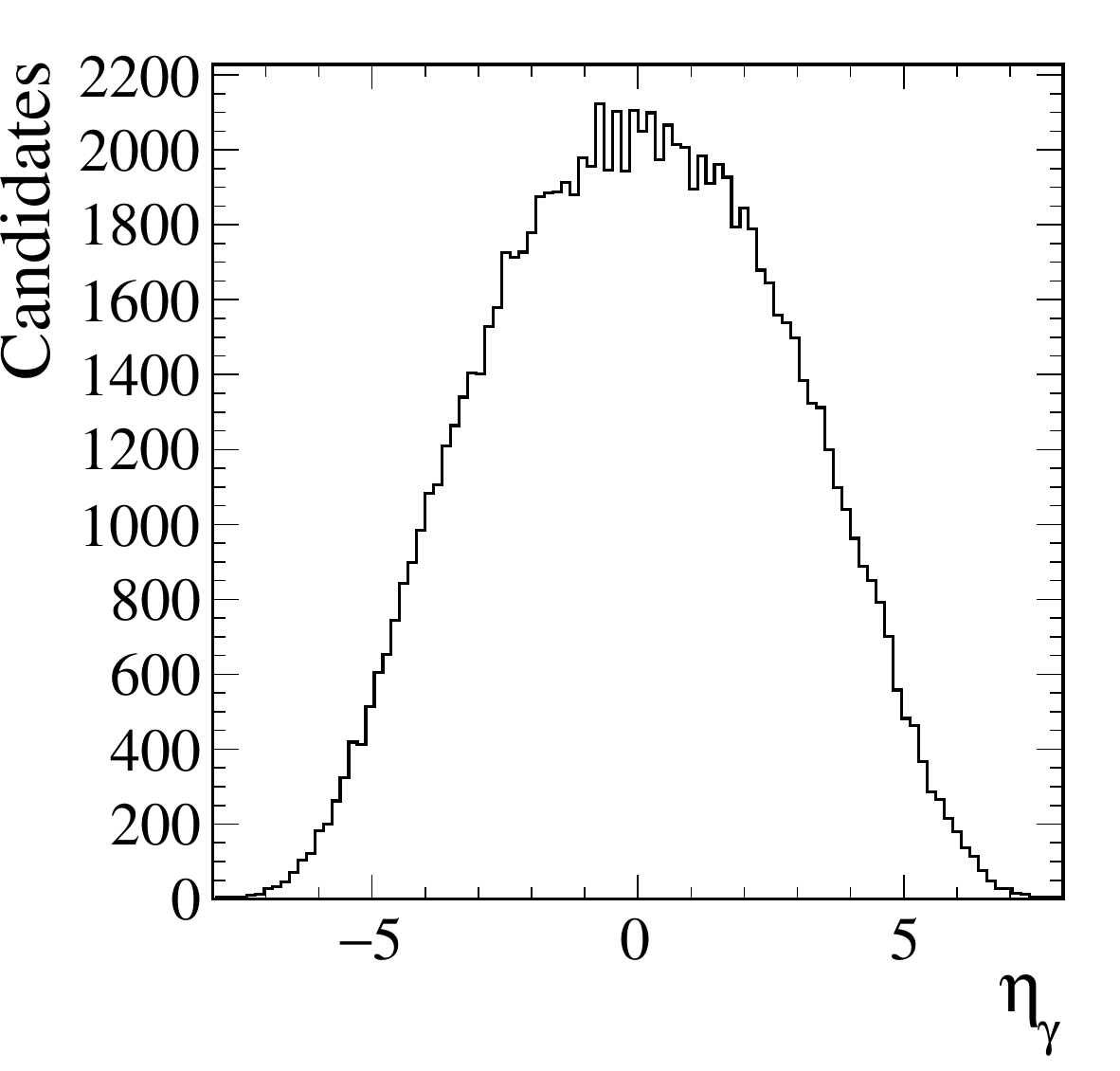} &
    \includegraphics[width=.33\textwidth]{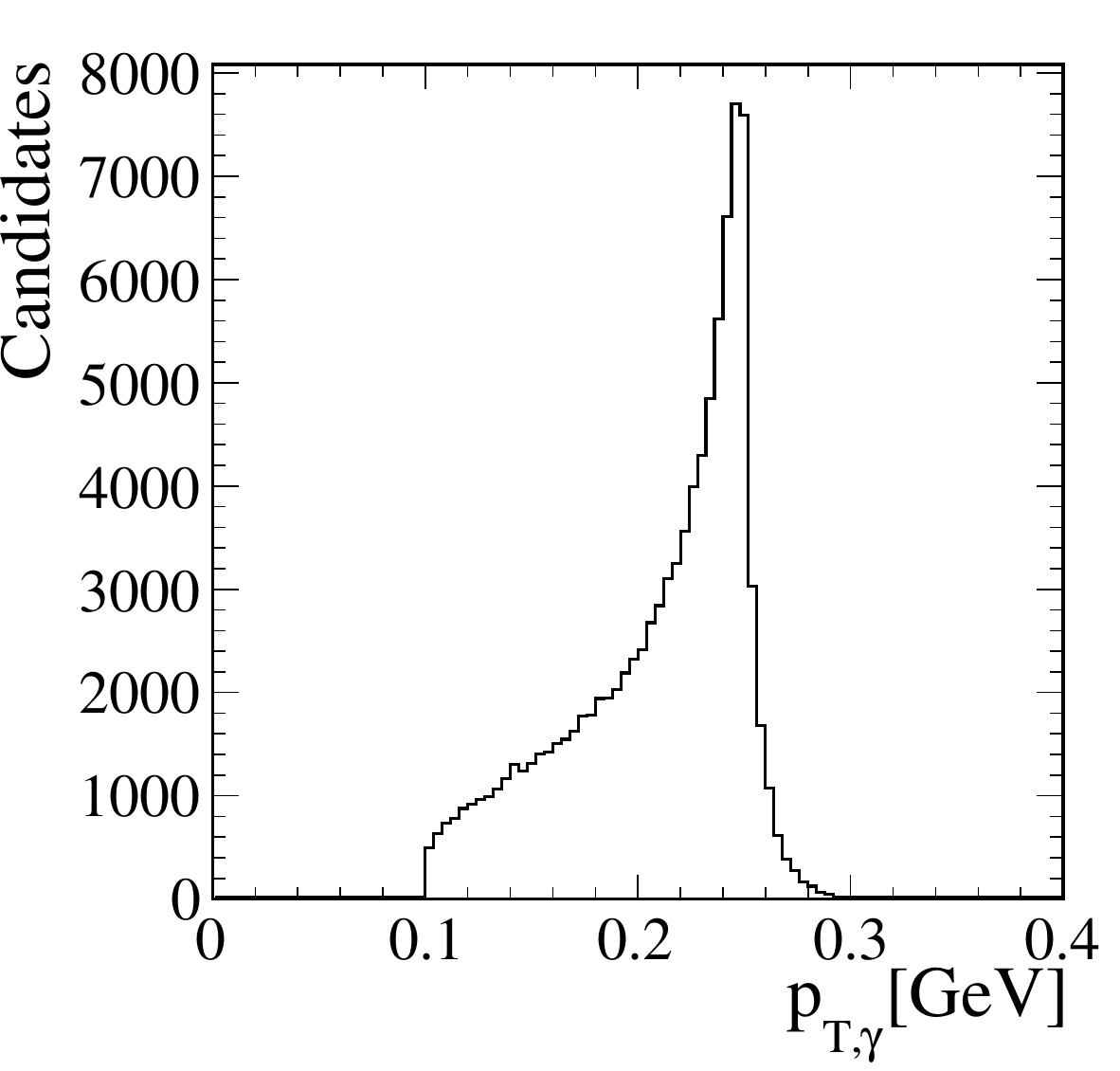} &
    \includegraphics[width=.33\textwidth]{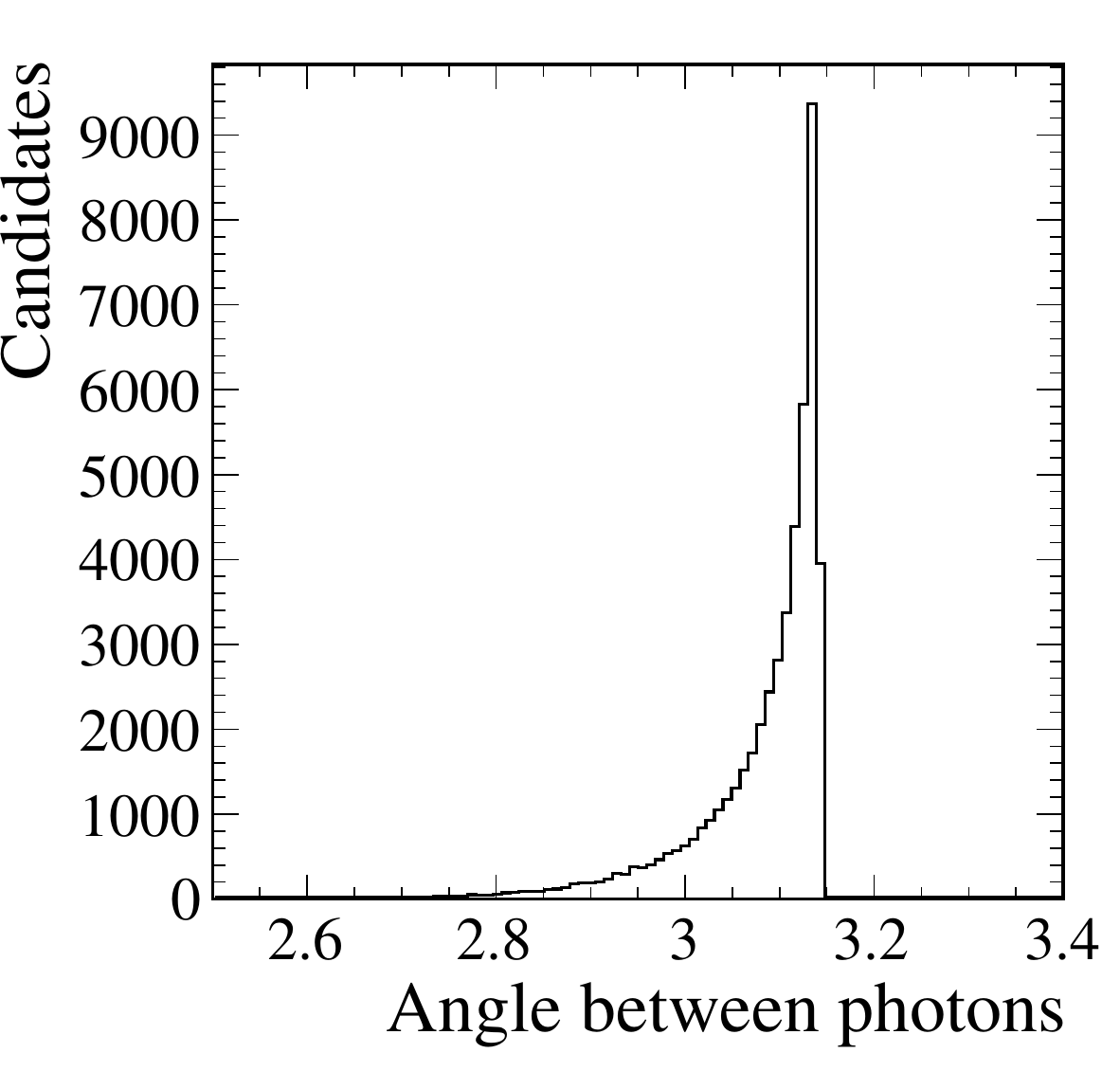}
    \end{tabular}
    \caption{Distributions of (left) pseudorapidity and (center) transverse momentum of the photons from ALP decays with $m_a=0.5$~GeV.  The azimuthal angle between the two photons from the ALP decay is shown on the right.}
    \label{photonnocuts}
\end{figure}

In order to estimate the ALP production in UPCs  with the experimental requirements described in the previous section, we will use the SuperChic v4.03 event generator~\cite{Harland-Lang:2020veo}. The ALP candidates will be generated in lead-lead collisions assuming a center of mass energy of $5.5$ TeV and that they have masses in the range of $0.2-1$~GeV. Following Ref.\cite{Harland-Lang:2020veo}, we will assume that the photon spectrum can be expressed in terms of the electric form factor  and that the absorptive corrections $S^2_{abs}({\mathbf b})$ for $\gamma \gamma$ interactions  can be estimated taking into account the multiple scatterings between the nucleons of the incident nuclei, which allows us to calculate the probability for no additional  ion -- ion rescattering at different impact parameters. The  pseudorapidity and transverse momentum of an ALP candidate with $m_a = 0.5$~GeV is shown in Fig.~\ref{paxion}. One has that the pseudorapidity distribution is larger for central rapidities and the transverse momentum distribution peaks at small $p_{T,a}$, which is expected since the incoming photons, emitted by the lead ions, are characterized by very small transverse momenta \cite{Baltz:2007kq}. 
Taking into account the ALP decay into a two - photon system, one has that the resulting photons will also be mainly produced at central rapidities  and transverse momentum of the order of 0.22 GeV, as demonstrated in the left and central panels of  Fig.  \ref{photonnocuts}. In addition, the result presented in the left panel, indicate that the  photons will be  back-to-back in the transverse plane. 




\begin{figure}[t]
    \centering
    \begin{tabular}{cc}
    \includegraphics[width=.48\textwidth]{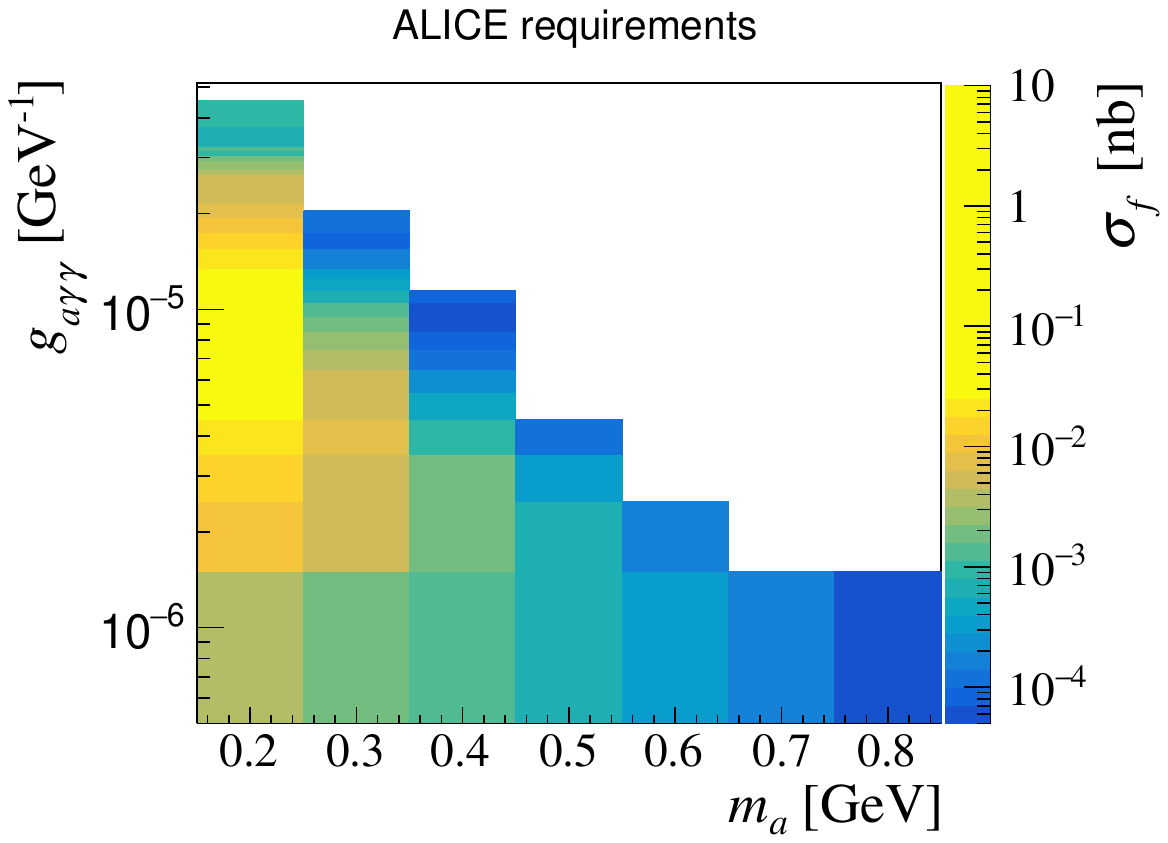} &   \includegraphics[width=.48\textwidth]{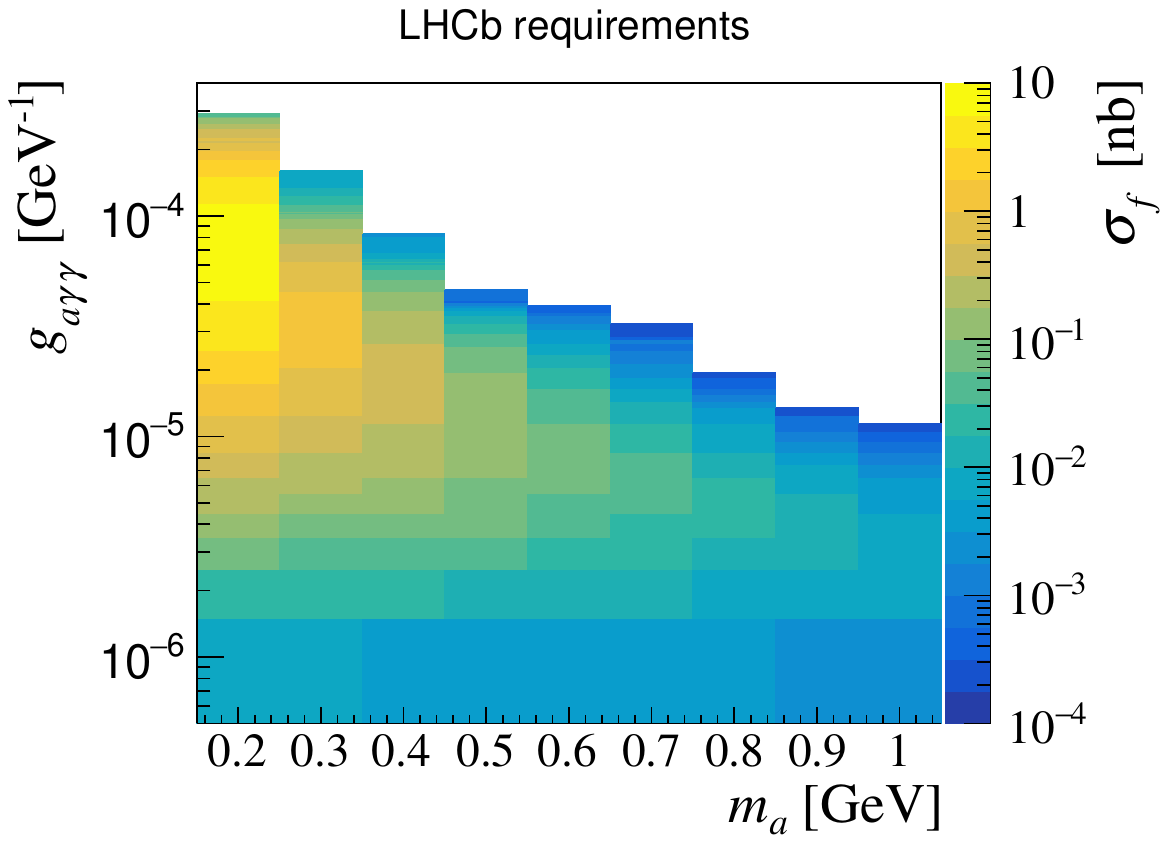} 
    \end{tabular}
    \caption{Fiducial cross-sections for ALP production in UPCs for different photon-ALP couplings and ALP masses. The predictions include the acceptance requirements for the photon pseudorapidity, the secondary vertex selection and probability of photon conversion considering the characteristics of the ALICE  (left panel) and LHCb (right panel) experiments.}
    \label{AlpFiducial}
\end{figure}

In what follows, we will present the results derived assuming the proposed search strategy. In order to be more realistic, the decay length values  for the ALPs are generated with an exponential distribution $\text{exp}(-l_{decay})$, with  the $l_{decay}$ of the ALPs with different values of $m_a$ being calculated using Eq.~(\ref{ldecay}). 
Moreover, we impose that the corresponding values must be larger than the PbPb luminous region in the transverse plane associated with the LHCb and ALICE experiments, which are assumed in  this study  to be $0.013$~cm~\cite{Aaij:2014zzy} for the LHCb experiment and $0.5$~cm~\cite{ALICE:2014sbx} for the ALICE experiment. It is important to emphasize that the background associated with the decay of light mesons is strongly reduced by this assumption, since these meson decay inside the luminosity region. The candidates selected with the experimental acceptance requirements are used to define 
fiducial cross-section $\sigma_{f}$ as follows:
\begin{equation}
\sigma_{f} = \sigma\frac{N_{\text{gen}}}{N_{\text{sel}}},
\end{equation}
where $\sigma$ is the ALP cross-section calculated by the event generator SuperChic~v4.03~\cite{Harland-Lang:2020veo}, $N_{\text{gen}}$ is the generated number of candidates and $N_{\text{sel}}$ is the number of ALP candidates after the requirements are applied. The corresponding results for the ALICE and LHCb experiments are presented in Fig.~\ref{AlpFiducial} considering different values for the ALP - photon coupling and ALP mass. One has that the cross-section is larger for smaller values of $m_a$, with the maximum value occurring for  $g_{a\gamma\gamma}$ of the order of $10^{-5}$ ($10^{-4}$)  GeV$^{-1}$ at the ALICE (LHCb) detector. Moreover, the values predicted for the LHCb detector are approximately one order of magnitude larger than for ALICE, and a larger range of the parameter space is covered by the LHCb detector. Such results are directly associated with the smaller value for the luminous region in this detector. Given the values for the cross - sections after the implementation of the proposed strategy, we can estimate the corresponding number of events for both experiments.
In 2018, the ALICE experiment collected 1~$\textrm{nb}^{-1}$ of integrated luminosity for heavy ion collisions~\cite{ALICE:2022xir}. As a consequence, the  expected number of ALPs decaying into two - photons is on the order of 10$^{-3}$ events. In contrast, The LHCb experiment collected 0.25~$\textrm{nb}^{-1}$ of integrated luminosity in 2018~\cite{LHCb:2022ahs}, which also suggests that the sensitivity for this search strategy can only be achieved with approximately $10^{3}$ more data. Such results indicate that, although the proposed search strategy eliminates the background associated with the decay of light mesons, the current $PbPb$ luminosities at the ALICE and LHCb detector still are small to allow us to use it to probing the axion - like particles.

\section{Summary}
\label{sec:sum}

Probing low mass axion - like particles decaying into two photons remains an experimental challenging due to the large background associated with decay of light SM mesons. 
Such a challenge  motivates the investigation of new searching strategies. 
In this paper, we have considered the possibility of eliminating the background by selecting events where the ALP decays outside the luminous region and the photons are  converted into electron-positron pairs inside the tracking detector. 
We have estimated the cross - section for ultraperipheral $PbPb$ collisions  at $5.5$ TeV and have taken into account the current characteristics of the ALICE and LHCb detectors in the implementation of the proposed searching strategy. 
Our results indicate that the LHCb detector covers a larger range of values for the ALP mass and ALP - photon coupling, which is covered in previous studies. 
However, the current experiments data sizes imply a very small number of events, which making this strategy unfeasible at present. 
Nevertheless, it is important to emphasize that both experiments collected PbPb collisions in 2023, and more data is expected in the coming years~\cite{Bruce:2021hii,dEnterria:2022sut}. 
While the predictions of this paper have a limited impact on current or near-future experiments, they present the first results for this novel ALP search strategy using secondary vertex selection in UPC.


\begin{acknowledgments}
 V.P.G. was partially supported by CNPq, FAPERGS and INCT-FNA (Process No. 464898/2014-5). M.S.R. and P.N. were partially supported by CNPq, FAPERJ and INCT-CERN (Process No. 406672/2022-9).

\end{acknowledgments}



\end{document}